\documentstyle[12pt]{article}
    \def\lsim{\raise0.3ex\hbox{$<$\kern-0.75em\raise-1.1ex\hbox{$\sim$}}}
\def\gsim{\raise0.3ex\hbox{$>$\kern-0.75em\raise-1.1ex\hbox{$\sim$}}}
\def\noi{\noindent}
\def\nn{\nonumber}
\def\bea{\begin{eqnarray}}  \def\eea{\end{eqnarray}}
\def\beq{\begin{equation}}   \def\eeq{\end{equation}}

\def\ben{\begin{enumerate}}  \def\een{\end{enumerate}}
   \def\cite#1{[\ref{#1}]}

\begin{document}
\begin{center}
\vbox to 1 truecm {}
{\bf COHERENCE EFFECTS IN CHARMONIUM PRODUCTION OFF NUCLEI : CONSEQUENCES FOR J/$\psi$ SUPPRESSION} \\

\vskip 8 truemm
{\bf A. Capella}\\
\vskip 5 truemm

{\it Laboratoire de Physique Th\'eorique\footnote{Unit\'e Mixte de
Recherche UMR n$^{\circ}$ 8627 - CNRS}
\\ Universit\'e de Paris XI, B\^atiment 210,
F-91405 Orsay Cedex, France} 
\end{center}

\vskip 1 truecm

\begin{abstract}
The probabilistic Glauber formula for nuclear absorption used in the literature is only valid at low energies and $x_+ \simeq 0$. Due to energy
conservation, $\sigma_{abs}$ is replaced by an effective cross-section $\sigma_{abs} + x_+^{\gamma}$ ($\sigma_{c\bar{c}-N} - \sigma_{abs})$ which increases
with $x_+$ and tends to the total $c\bar{c}-N$ cross-section $\sigma_{c\bar{c}-N}$. Experimental data can be described with $\sigma_{abs} \sim 4 \div 5$~mb
and $\sigma_{c\bar{c}-N} \sim 15 \div 20$~mb. At high energies, due to the increase of the coherence length, this formula changes. The main change is the
replacement of $\sigma_{abs}$ by $\sigma_{c\bar{c}-N}$ -- for all values of $x_+$, as $s \to \infty$. Thus, if $\sigma_{c\bar{c}-N} > \sigma_{abs}$ the
$J/\psi$ suppression due to nuclear interaction will increase with energy.
\end{abstract}

\vskip 1 truecm

\noi LPT Orsay 02-63\par
\noi June 2002\par
\par \vskip 1 truecm

\noindent {\bf Talk presented at the International Workshop on Charm Production, ECT*, Trento, June 2002}
\newpage
\pagestyle{plain}
\noi \underbar{\bf LOW ENERGY PROBABILISTIC FORMULA.} Let us consider a proton-nucleus collision. The probabilistic Glauber formula for the $J/\psi$
survival is well known. A $c\bar{c}$ pre-resonant state is produced inside the nucleus at some value $Z$ of the longitudinal coordinate, and interacts with
the nuclei on its path. If the interaction is inelastic, the $c\bar{c}$ pair can transform into another pair which has vanishing projection into $J/\psi$.
The corresponding cross-section is called 	absorptive cross-section $\sigma_{abs}$. We can write $\sigma_{abs} = (1 - \varepsilon)\sigma_{c\bar{c}-N}$,
where $\sigma_{c\bar{c}-N}$ is the total $c\bar{c}-N$ cross-section and $\varepsilon \sigma_{c\bar{c}-N}$ is the contribution to $\sigma_{c\bar{c}-N}$ of
those intermediate states which have a non-zero projection into $J/\psi$. In the Glauber model one has

\bea 
\label{1e}
\sigma_{pA}^{\psi}(b) &=& \sigma_{pN}^{\psi} \ A \int_{-\infty}^{+ \infty} dZ \ \rho_A (b, Z) \exp \left [ - \sigma_{abs} \ A \int_Z^{\infty} dZ' \ \rho_A (b,
Z')\right ] \nn \\
&=& \left ( \sigma_{pN}^{\psi}/\sigma_{abs}\right ) \left [ 1 - \exp \left ( - \sigma_{abs} \ A\ T_A(b) \right ) \right ] \ .\eea

\noi Note that for open charm production $\sigma_{abs} = 0$ and $\sigma_{pA} = A \sigma_{pN}$. \\

\noi \underbar{\bf ASYMPTOTIC FIELD THEORY FORMULA.} At high energies the coherence length increases and the projectile interacts with the nucleus as a
whole \cite{2r}. Thus, it is no longer possible to consider its collisions as ``successive'' and expression (\ref{1e}) is changed. As we shall see below
this change consists in the replacement \cite{1r} 

\beq
\label{2e}
\left ( 1/\sigma_{abs}\right ) \left [ 1 - \exp \left ( - \sigma_{abs}\ A \ T_A(b) \right ) \right ] \Rightarrow A\ T_A(b) \exp \left [ - {1 \over 2}\ 
\widetilde{\sigma}\ A \ T_A(b) \right ] \eeq

\noi where $\widetilde{\sigma}\equiv \sigma_{c\bar{c}-N}$. The change is twofold. There is a change in the form of the expression and $\sigma_{abs}$ has
been replaced by $\sigma_{c\bar{c}-N}$. If $\sigma_{abs} \sim \sigma_{c\bar{c}-N}$ the change (\ref{2e}) is not important. Indeed the two expressions
coincide at the first and second order in $\widetilde{\sigma}$ and, since $\sigma_{abs}$ is not large, the result will not be significantly changed.
However, if $\sigma_{c\bar{c}-N} \gg \sigma_{abs}$, there will be a very important increase of the so-called normal $J/\psi$ suppression (which can no
longer be called nuclear absorption since it is due to the total cross-section $c\bar{c}-N$, rather than to its absorptive part). This can be the case if
the $c\bar{c}$ pair is produced in a color state and is accompanied by light quarks in order to have a colorless system. In this case,
$\sigma_{c\bar{c}-N}$ can be a typical hadronic cross-section, due to the presence of the light quarks \cite{3r}. Furthermore, the probability for this
$c\bar{c}$ system to keep its projection into $J/\psi$ after interacting inelastically has to be large, i.e. $\varepsilon \approx 1$. \par

The derivation of the asymptotic formula is straightforward. Let us denote by $\sigma$ the cross-section for the interaction of the light quarks in the
projectile and $\widetilde{\sigma} \equiv \sigma_{c\bar{c}-N}$ that of the produced $c\bar{c}$ system (we assume, for simplicity, that the corresponding
amplitudes are purely imaginary). Let us single out one of the light quark interactions in which the $c\bar{c}$ pair is produced (for instance via
gluon-gluon fusion). We have \cite{1r}

\beq \label{3e} 
\sigma_{pA}^{\psi}(b) = \sigma_{pN}^{\psi} \left [ \sigma_A (\sigma + \widetilde{\sigma}) - \sigma_A (\widetilde{\sigma}\right ] / \sigma
\eeq

\noi where

\beq \label{4e} 
\sigma_A(\sigma) = 2 \left \{ 1 - \exp \left [ - {1\over 2} \ \sigma\ A \ T_A(b) \right ] \right \}
\eeq

\noi is the Glauber total cross-section in terms of the cross-section $\sigma$ for the scattering on a single nucleon. In eq. (\ref{3e}) we have
substracted the term with no light quark interaction (since we assume that at least one such interaction is needed to produce the $c\bar{c}$ pair). The
division by $\sigma$ is due to the fact that one such collision is precisely the one producing the $c\bar{c}$-pair -- first factor of (\ref{3e}). From
(\ref{3e}) and (\ref{4e}) we get  

\beq \label{5e} 
\sigma_{pA}^{\psi}(b) = \left ( \sigma_{pN}^{\psi}/\sigma \right ) \sigma_A(\sigma ) \exp \left  [ - {1 \over 2} \ \widetilde{\sigma} \ A\ T_A(b) \right ]
\ . \eeq

\noi We see that the terms containing $\sigma$ and $\widetilde{\sigma}$ factorize. The former produce nuclear effects (shadowing) in the nucleus vertex
function (which is no longer proportional to $A$). The terms containing $\widetilde{\sigma}$ correspond to the rescattering of the $c\bar{c}$ system and
have the form given by (\ref{2e}). \par

As stated above, the main change between the probabilistic formula and the asymptotic one consists in the replacement of $\sigma_{abs}$ by
$\sigma_{c\bar{c}-N}$. This change is entirely due to coherence effects. At low energy, some contributions vanish due to the fact that there is a
non-vanishing minimum momentum transfer $(t_{min} \not= 0$) -- and this contribution is suppressed by the nuclear form factor. It turns out \cite{3r}
that the other contributions cancel with each other, giving rise to an $A^1$ behaviour typical of open charm production. The only case in which this
cancellation is avoided is when the $c\bar{c}$ pair is ``destroyed'', (i.e. converted into a $c\bar{c}$ system with vanishing projection on the
$J/\psi$). For this reason only the absorptive part of the total $c\bar{c}-N$ cross-section plays a role at low energy. At asymptotic energy $t_{min} =
0$. Thus, the above cancellation does not occur and the total $c\bar{c}-N$ cross-section contributes to the $J/\psi$ suppression.\par

The expression of the shadowing corrections in eq. (\ref{5e}) is not realistic since the Glauber expression  has been used. It is, indeed, well known that
shadowing corrections have to be described by triple Pomeron diagrams. In view of that, in the following we shall only consider in eq. (\ref{3e}) the
terms linear in $\sigma$, and will use a standard formulation of the shadowing corrections (which correspond to higher order terms in $\sigma$). It is,
nevertheless, instructive to see from the complete expression (\ref{3e}) that the shadowing corrections vanish in the low energy limit and are
factorizable in the high energy one. In the following we will assume that factorization also holds at intermediate energies -- and will denote the
corresponding cross-section $\sigma_{pN}^{\psi, shadow}$. \\

\noi \underbar{\bf INTERPOLATING FORMULA.} Using the AGK cutting rules \cite{4r} it is possible to obtain the exact formula at finite energies. We have
\cite{1r} instead of (\ref{5e}) 
   
\beq \label{6e} 
\sigma_{pA}^{\sigma}(b) = \left ( \sigma_{pN}^{\psi , shadow}/\sigma \right ) \sum_{n=1}^A {A \choose n} \sum_{j=1}^n \ T_n^{(j)}\ \sigma_n^{(j)} 
\eeq

\noi where

\bea \label{7e} 
&&\sigma_n^{(j)} (b) = \sigma \left ( - {\widetilde{\sigma} \over 2} \right )^{j-1} \left [ - (1 - \varepsilon ) \widetilde{\sigma}\right ]^{n-j} - \nn \\
&&\sigma \left ( - {\widetilde{\sigma} \over 2} \right )^{j-1} (j - 1) \left [ {\widetilde{\sigma} \over 2} - (1 - \varepsilon ) \widetilde{\sigma}
\right ] \left [ - (1 - \varepsilon ) \widetilde{\sigma} \right ]^{n-j} = \nn \\
&&\left ( - {\widetilde{\sigma} \over 2} \right )^{j-1} \ j \left [ - (1 - \varepsilon ) \widetilde{\sigma} \right ]^{n-j} - \left ( - {\widetilde{\sigma}
\over 2} \right )^{j-2}\ (j-1) \left [ - (1 - \varepsilon ) \widetilde{\sigma} \right ]^{n-j+1} \eea

\noi and

\beq \label{8e} 
T_n^{(j)}(b) = n ! \int_{-\infty}^{+\infty} dZ_1 \int_{Z_1}^{+ \infty} dZ_2 \cdots \int_{Z_{n-1}}^{+ \infty} dZ_n \cos (\Delta (Z_1 - Z_j) \prod_{i=1}^n
\rho_A (b, Z_i) \ . \eeq

\noi Here $\Delta$ is the inverse of the coherence length 

\beq \label{9e} 
\Delta \equiv {1 \over \ell_c} = m_p \ {M_{c\bar{c}}^2 \over s \ x_1}
\eeq

\noi with $x_F = x_1 - x_2$ and $x_1 x_2 s = M_{c\bar{c}}$. \par

In (\ref{8e}) the longitudinal coordinates of the interacting nucleons have been ordered as $Z_1 \leq Z_2 \cdots \leq Z_n$. The index $j$ in (\ref{7e}) and
(\ref{8e}) denotes the first interaction (i.e. the one with the smallest value of $Z$) which has been cut in such a way that $t_{min} = 0$ for the $n - j$
ones with $Z > Z_j$. For details see ref. \cite{1r}. The damping factor $\cos (\Delta (Z_1 - Z_j))$ depends on the difference $Z_1 - Z_j$. The only case
in which $t_{min} = 0$ is when $j = 1$, i.e. when the first interaction is cut. In all other cases $t_{min} \not= 0$. Therefore, the term with $j = 1$ is
the only one that survives in the limit $s \to 0(\Delta \to \infty )$. It is easy to verify that one recovers in this way the probabilistic formula
(\ref{1e}) -- which depends only on $\sigma_{abs}$. Likewise, it can be verified\footnote{For $\Delta  = 0$, $T_n^{(j)} \equiv T_A^n$. Moreover, if one
considers the two terms in the last equality of (\ref{7e}), there is a cancellation between the first term of $\sigma_n^{(j)}$ and the second term of
$\sigma_n^{(j+1)}$. The only term left is thus the first term of $\sigma_n^{(n)}$.} that, in the limit $s \to \infty (\Delta \to 0)$ we recover the
asymptotic expression (\ref{5e}) -- which depends only on $\sigma_{c\bar{c}-N}$.\par

We see from (\ref{9e}) that at fixed $s$ and $x_2$, $\Delta$ increases with $x_1$. Therefore, if $\sigma_{c\bar{c}-N} > \sigma_{abs}$, $J/\psi$ suppression
will increase with $x_1$ -- a feature observed experimentally. In ref. \cite{5r} an attempt was made to describe the data in this framework. However, as
recognized in \cite{5r}, there is a caveat. Since $\Delta$ is a function of $x_2$, one obtains a scaling in $x_2$, whereas the data indicate rather a
scaling in $x_1$ (or $x_F$). A way out is to take into account the modifications of the AGK rules resulting from energy conservation \cite{3r}. Indeed,
by energy conservation, the $x$ distribution of $J/\psi$ gets softer when the number $K$ of inelastic collisions increases. This has been taken
into account in \cite{3r} introducing a factor  

\beq \label{10e} 
F_k (x_1) = \left  (1 - x_1^{\gamma} \right )^k
\eeq

\noi in the contribution to $\sigma_{pA}^{\psi}$ corresponding to $k$ inelastic collisions of the $c\bar{c}$ system. In ref. \cite{3r} this has been done
in the probabilistic Glauber model. In order to do it in the general framework described above, one has to decompose $\sigma_n^{(j)}$, eq. (\ref{7e}),
into a sum of terms $\sigma_{n,k}^j$ corresponding to a fixed number $k$ of inelastic collisions. It turns out that this amounts to replacing $[-(1-
\varepsilon )\widetilde{\sigma}]^{n-j}$ by $\sum\limits_{k=0}^{n-j} {n- j \choose k} (\varepsilon \widetilde{\sigma})^k (- \widetilde{\sigma})^{n-j-k}
F_k(x_1)$. With the form of $F_k$ in eq. (\ref{10e}) the summation in $k$ can be performed analytically. The final result is

\beq \label{11e} 
\sigma_n^{(j)} (b) = \left ( - {\widetilde{\sigma} \over 2} \right )^{j-1}\ j\ \left [ -\sigma_{eff}\right ]^{n-j} - \left ( - {\widetilde{\sigma} \over 2} \right
)^{j-2} \ (j-1) \left [ - \sigma_{eff} \right ]^{n-j+1}\eeq

\noi where

\beq \label{12e} 
\sigma_{eff} = \widetilde{\sigma} (1 - \varepsilon ) + \widetilde{\sigma} \varepsilon \ x_1^{\gamma} \ .
\eeq

Thus, as a consequence of the modification of the AGK rules due to energy conservation, we have obtained the same expression (\ref{7e}) with
the absorptive cross-section $\sigma_{abs} = (1 - \varepsilon ) \widetilde{\sigma}$ replaced by $\sigma_{eff}$. Therefore, the asymptotic
limit, eq. (\ref{5e}), which does not depend on $\sigma_{abs}$ is not changed. \par

As discussed above, in the low energy limit, $\Delta \to \infty$, only the term $j = 1$ survives. We obtain in this case 

\beq \label{13e} 
\sigma_{pA}^{\psi} (b) = \left ( \sigma_{pN}^{\psi}/\sigma_{eff} \right ) \left [ 1 - \exp \left ( - \sigma_{eff} \ A\ T_A(b)\right ) \right ]
\ . \eeq 

\noi We see that for $x_1 \to 0$ we recover the probabilistic expression (\ref{1e}) with $\sigma_{eff} \sim \sigma_{abs}$. However, for $x_1
\to 1$ we have $\sigma_{eff} \sim \sigma_{c\bar{c}-N}$. While coherence effects produce this change from $\sigma_{abs}$ to 
$\sigma_{c\bar{c}-N}$ as $s \to \infty$, the introduction of the factors $F_k(x_1)$, eq. (\ref{10e}), leads to the same effect at low $s$ as
$x_1$ increases. We recover in this way scaling in $x_1$ at low energies. \par

As shown in ref. \cite{3r}, eqs. (\ref{12e})-(\ref{13e}) give a good description of experimental data with $\sigma_{abs} = 5$~mb,
$\sigma_{c\bar{c}-N} = 20$~mb ($\varepsilon = 0.75$) and $\gamma = 2$. It was also shown in \cite{3r} that $\gamma$ increases with the mass of
the heavy-quark pair and thus, the effect of energy conservation via the factors $F_k$ concentrates to larger values of $x_1$ when this mass
increases.\\

\noi \underbar{\bf CONCLUSIONS AND OUTLOOK.} We have shown that the probabilistic Glau\-ber formula for the survival of $J/\psi$ in a $pA$
collision changes with energy due to coherence effects. The main change is the replacement of $\sigma_{abs}$ by $\sigma_{c\bar{c}-N}$ -- the
total $cc-N$ cross-section. The same change occurs at low energy when $x_1$ increases from 0 to 1, due to energy conservation. If
$\sigma_{abs} \simeq  \sigma_{c\bar{c}-N}$ the discussion in this paper is of little phenomenological relevance. Nuclear absorption will be
similar at RHIC and SPS energies. Moreover, the increase of the $J/\psi$ suppression with $x_F$ cannot be due to the mechanism described here.
If, on the contrary, $\sigma_{c\bar{c}-N}$ is substantially larger than $\sigma_{abs}$, the increase with $x_F$ is well reproduced and a huge
increase of the $J/\psi$ suppression in $pA$ collisions at RHIC and LHC is predicted.\par

Numerical calculations using eq. (\ref{11e}) are in progress \cite{6r}. They should provide a more precise determination of the value of
$\sigma_{c\bar{c}-N}$ from available data. They should also allow to determine quantitatively how the $x_F$ scaling is converted into a
scaling in $x_2$ as energy increases -- first at low $x_F$ and then extending to larger values of $x_F$ as the energy increases.\\

\noi {\bf ACKNOWLEDGMENTS}\\

It is a pleasure to thank all my collaborators in the subject of this contribution~: N. Armesto, M. Braun, K. Boreskov, A. Kaidalov, C. Pajares, C.
Salgado and J. Tran Thanh Van. I also thank A. Polleri and the organizers of the ECT International Workshop on Charm Production for an interesting and
stimulating meeting.\\

\def\labelenumi{[\arabic{enumi}]}
\noindent
{\bf REFERENCES}

\ben
\item\label{1r} M. A. Braun, C. Pajares, C.A. Salgado, N. Armesto, A. Capella, Nucl. Phys. B {\bf 509}, 357 (1998).\\
M. A. Braun and A. Capella, Nucl. Phys. B {\bf 412}, 260 (1994).
 
\item\label{2r} B. Z. Kopeliovich, A. V. Tarasov and J. Huefner, hep-ph/0104256. 

\item\label{3r} K. Boreskov, A. Capella, A. Kaidalov and J. Tran Thanh Van, Phys. Rev. D {\bf 47}, 919 (1993).
 
\item\label{4r} V. Abramovsky, V. N. Gribov and O. V. Kancheli, Sov. J. Nucl. Phys. {\bf 18}, 308 (1974).
 
\item\label{5r} C. A. Salgado, hep-ph/0105231.

\item\label{6r} N. Armesto, A. Capella and C. A. Salgado, in preparation.
\een

\end{document}